\documentclass[prb,twocolumn,floatfix,showpacs,epsf,psfig]{revtex4-2}
\usepackage{epsfig}
\usepackage{graphicx}
\usepackage{dcolumn}
\usepackage{bm}
\usepackage{subfigure}
\usepackage{blindtext}
\usepackage{amsmath}

\begin{document}
\title{Electron pair emission from surfaces: Some general experimental considerations}
\author{R. Kamrla, W. Widdra}
\affiliation{Institute of Physics, Martin-Luther-Universit\"{a}t Halle-Wittenberg, Von-Danckelmann-Platz 3, D-06120 Halle (Saale), Germany}
\author{C.-T. Chiang}
\affiliation{Institute of Atomic and Molecular Sciences, Academia Sinica, Taipei, Taiwan}
\author{F. O. Schumann}
\altaffiliation{Electronic mail: schumann@mpi-halle.de}
\affiliation{ Max-Planck-Institut f\"{u}r Mikrostrukturphysik, Weinberg 2, 06120 Halle, Germany}
\date{\today}
\begin{abstract}
We discuss some experimental facets of electron pair emission from surfaces using two different experimental approaches. In the first case the instrument consists of a pair of hemispherical analyzers which are operated with continuous primary beams of electrons or photons. The second instrument employs a pair of time-of-flight spectrometers which require a pulsed excitation source. A key experimental quantity is the ratio of `true'\ to `random'\ coincidences which can be determined in different ways. Regardless of the type of instrument the primary flux has to adopt a much smaller value than in single electron spectroscopy. We describe different approaches to obtain the relevant count rates, in particular the concept of operating with a delayed coincidence circuit. We also address the question on how to compare the two types of spectrometer in terms of their performance.  
\end{abstract}
\maketitle
\section{Introduction}
\begin{figure}[t]
	\includegraphics[width=7.5cm]{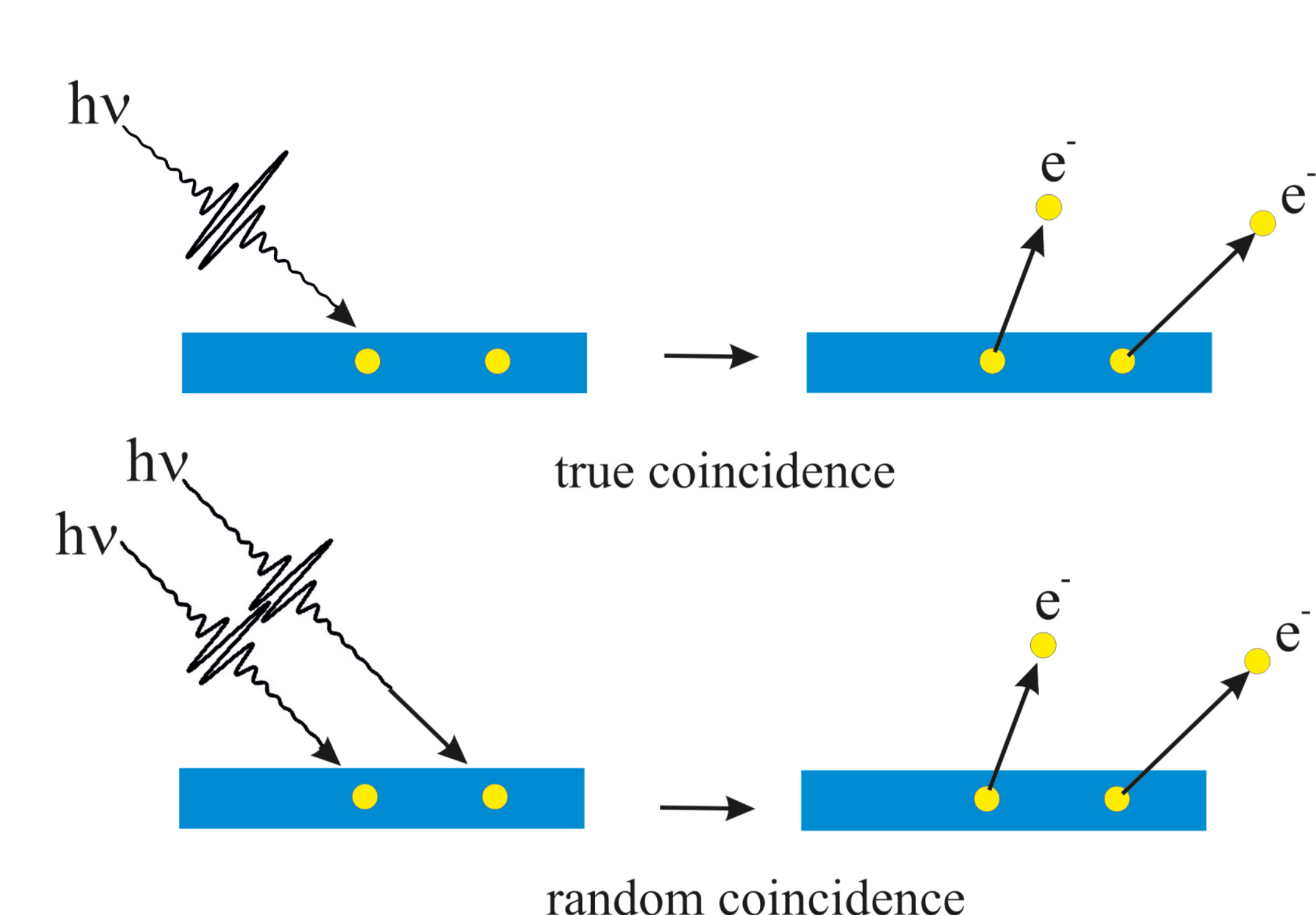}
	\caption{Sketch of the emission of an electron pair via single photon absorption (top) or absorption of two photons (bottom). The first case is the process of interest and leads to true coincidences. The other pathway leads to unwanted background commonly referred to as `random'\ coincidences. \label{true_random_cartoon}}
\end{figure}

In the past decades photoelectron spectroscopy has developed into the most powerful experimental technique to determine the detailed electronic properties of matter. On the experimental side, a wide range of energy-, momentum- and spin-resolving electron spectrometers has been employed. Here the large majority of spectrometers records single electron spectra, e.g. the kinetic energy distribution upon photon absorption. However, also the emission of two electrons excited by absorption of a 
single photon is possible, which might be due to a strong electron-electron correlation. This process can be identified if two time-resolving electron detector are used. Both detectors are connected via a coincidence logic. Only if both electrons hit the detectors within an appropriate small relative time window, the two-electron event is registered. 
In general, the coincidence rates are often rather small, in the range of 0.1 to 10 events/s. Experimentally it is challenging to discriminate these true coincidence events from the large number of single-electron events originating from different photons.\linebreak
\indent To achieve the experimentally desired signal-to-noise ratios, adequate event statistics are required. The resulting measurement time scales with the inverse of the repetition rate of the pulsed excitation. Therefore high repetition rates are favored, which are only limited  by the requirement that the slowest electrons have reached the detector within one excitation periode. The latter is typically in the order of a few MHz and is fulfilled at  synchrotron light sources in special operation modes as well as with pulsed UV laser sources.
In Fig.\ref{true_random_cartoon} electron pair emission due to single-photon absorption  known as double photoemission (DPE) is illustrated.  An unwanted background signal arises, if two photons are independently absorbed as this will also lead to the emission of an electron pair. It is customary to refer to the latter process as `random'\ coincidences whereas the genuine DPE signal is known as `true'\ coincidence.
As we discuss later the process of interest scales linearly with the primary flux, while the undesirable process scales quadratically with the flux. This unwanted pathway very quickly overwhelms the genuine signal and the operation has to commence with much reduced flux.
Let us quote a numerical example for electron pair emission due to primary electron impact, known as (e,2e). If one utilizes  an experiment  with energy dispersive elements, as described later,  the sample current is $10^{-13}$ A.  This is many orders of magnitudes lower than for Auger electron spectroscopy using a primary electron beam for which primary currents are of the order of a few $\mu$A.
The requirement of low primary flux asks for a detection scheme with large detection probability. This means that the detector solid angle should be as large as possible while a large electron kinetic energy window should be covered. There exist different solutions for experiments in the gas phase or at surfaces.\cite{1098_Jensen,1039_Thurgate,889_Herrmann,1038_Gotter,1619_Ullrich,1006_Hattass, 1712_Eland,2334_Penent}\linebreak
\indent These schemes can be separated into energy dispersive detection or time-of-flight (ToF) spectroscopy. The latter asks for a pulsed excitation source which in the case of photon beams is more involved. For efficient measurement scheme a repetition rate of the order of a few MHz is needed, because once the slowest electron has been detected the next excitation can commence. This requirement can be fulfilled at synchrotron  light sources in special operation modes, but the available beamtime is restricted.\linebreak
%
%
%
\indent More recent developments in laser technology using high-order harmonic generation (HHG) allow to perform in-house DPE experiments.\cite{1794_Chiang,1865_Chiang,1932_Chiang,2268_Chiang,2111_Truetzschler} While it is possible to carry out these type of experiments also with an energy dispersive element together with a standard vacuum-ultraviolet VUV light source, the available photon energies are limited to a few lines. These sources employ the discharge of nobel gases of He, Ar and  Ne.\cite{1833_Schoenhense} We use in our work  a monochromatized He light source.\cite{MBS_lamp,Scienta_mono} For ToF spectroscopy,  HHG light sources provide more options.\linebreak
\indent It is well-established on how to determine the individual contribution of `random'\ and `true'\ coincidences. Here  we  discuss  a novel approach for ToF-spectroscopy similar to a  recently reported concept in Auger-photoelectron coincidences exploring synchrotron radiation on surfaces, for a recent overview about coincidence spectroscopy we refer to the literature.\cite{2325_Leitner,1938_Arion}
%
%
\indent In this work we  compare this with the common approach of a double logarithmic analysis. We will demonstrate that both approaches yield the same characteristic parameters for different DPE experiments in which the intensity levels are orders of magnitude different.\linebreak
 \begin{figure}[b]
	\includegraphics[width=7.5cm]{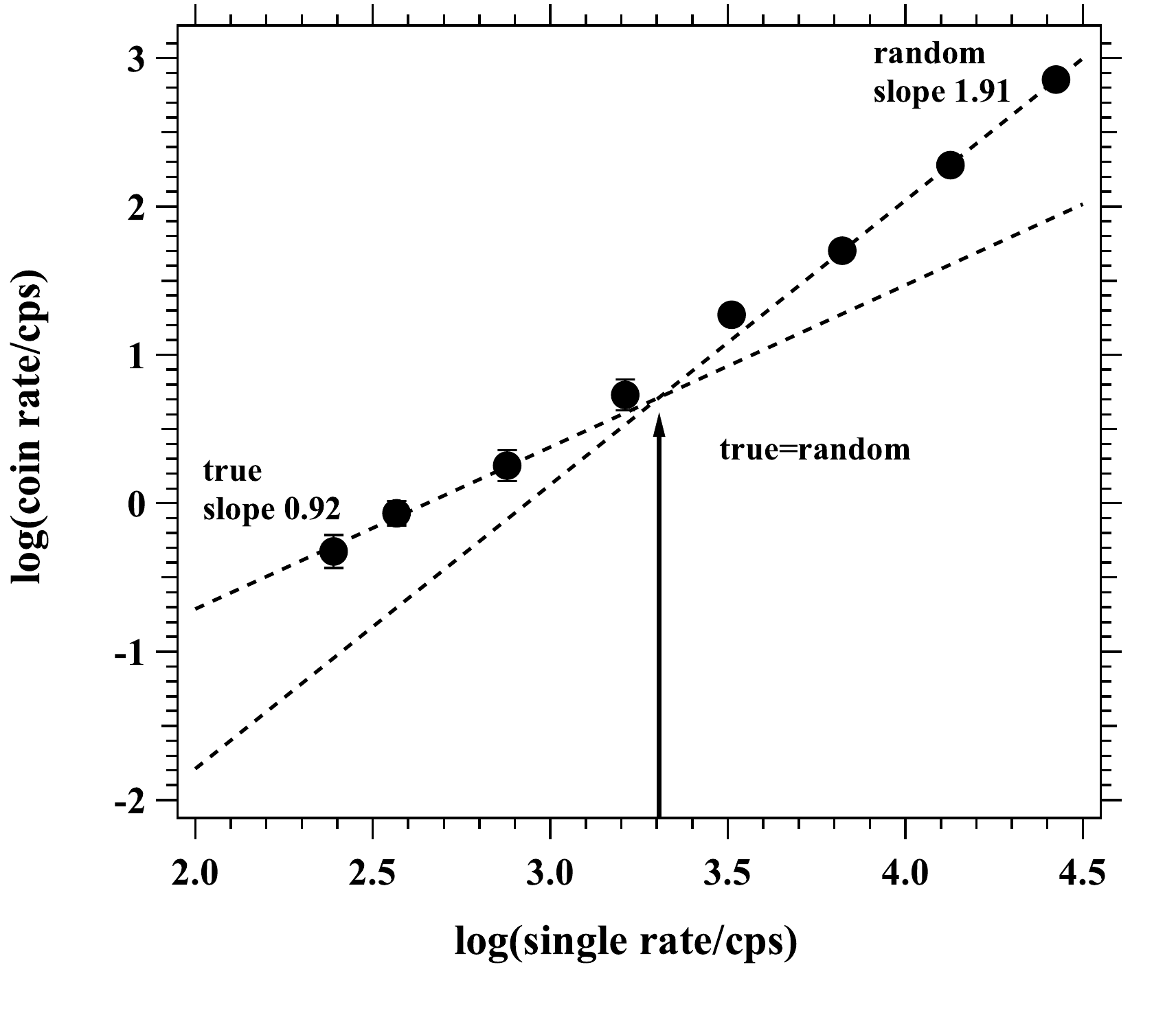}
	\caption{ We display an example of the sum of `true'\ and `random'\ coincidence count rates as a function of single photoemission count rate in a double logarithmic manner. The arrow indicates the singles rate for which the number of `true'\ and `random'\ coincidences are equal.\label{log_log_scienta}}
\end{figure}
\indent We start by considering how a genuine pair emission can be identified and how the statistical probabilities for `true'\ and `random'\ events have to be formulated. Second, we introduce the concept of delayed coincidences and how the relevant quantities can be experimentally determined. Third, we provide experimental data to prove the concept and how it can be related to a known procedure. Finally, we provide an outlook on how the `random'\ events can be removed from coincidence event spectra.
\section{signature of `true'\ coincidences}
\begin{figure}[b]
	\includegraphics[width=8.0cm]{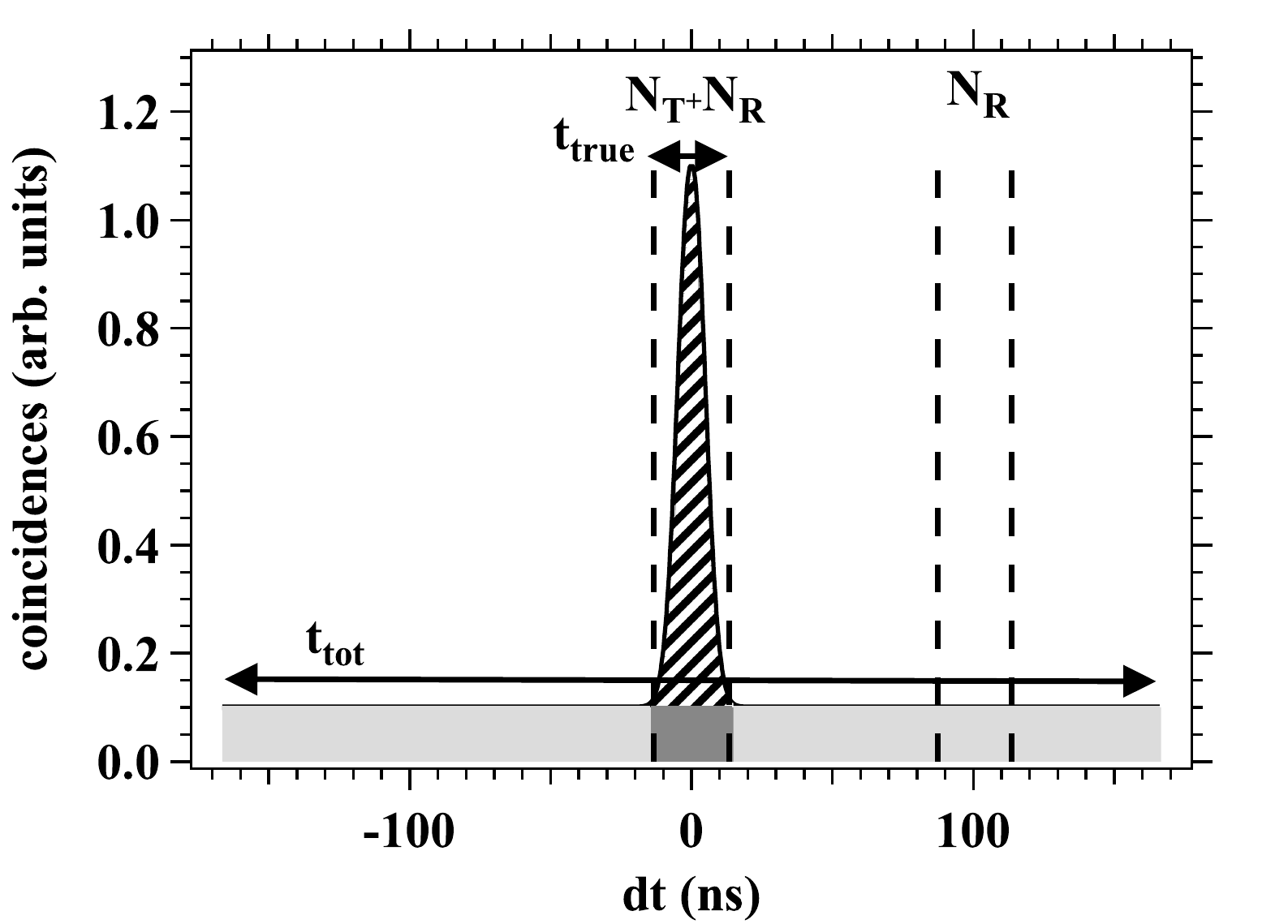}
	\caption{Sketch of the arrival time histogram. The emergence of the peak is proof of the emission of electron pairs. The width $t_{true}$ is a consequence of the time dispersion of the spectrometer. The histogram has a total width of $t_{tot}$.\label{hist_cartoon}}
\end{figure}

Although our discussion will be valid for a variety of experimental approaches, we will present data obtained from two different set-ups for the investigation of solid surfaces. Our ToF instrument consists of a pair of lenses. The lens axes include an angle of 90$^\circ$ and have an angular opening of $\pm$15$^\circ$.\cite{1804_Huth} As excitation sources a pulsed electron gun or an HHG light source is available. The light source can operate at a repetition rate of 0.2-1 MHz and provides a  photon energy range from 13 to 39~eV.\cite{1865_Chiang,2111_Truetzschler,2268_Chiang} The energy dispersive experiment utilizes a pair of hemispherical analyzers.\cite{1588_Schumann} Also in this case the two electron-optical axes include an angle  90$^\circ$ with an angular acceptance   of $\pm$15$^\circ$ within the scattering plane.
This instrument has proven to be very versatile. We reported on experiments with primary electrons, positrons, He$^{2e+}$ ion and photons from laboratory and synchrotron sources.\cite{1825_Brandt,1853_Wei,2019_DiFilippo,2113_Li,2250_Schumann}\linebreak
\indent Let us define the probability to find one photon in a light pulse by $P_1(\lambda)$ while the probability $P_2(\lambda)$ describes the probability to find two photons in a pulse. The quantity $\lambda$ is the average number of photons in a pulse and hence proportional to the primary flux. If we assume a Poisson distribution and $\lambda\ll 1$ we obtain the following expressions:
\begin{equation}  P_{1}(\lambda)=\lambda~~,   \\  ~~ P_{2}(\lambda)=\frac{1}{2}\lambda^2   
	\label{poisson1}
\end{equation}
The key result is that the probability  $P_{2}(\lambda)$ goes quadratically with $\lambda$ and therefore with the flux, while the term $P_{1}(\lambda)$ displays a linear relation. This explains the flux dependence for `random'\ and  `true'\ coincidences, respectively.\linebreak
\indent This difference in the flux dependence can be observed by presenting the coincidence rate as a function of singles rate in a double logarithmic manner, see Fig.\ref{log_log_scienta}. The dashed lines indicate the behavior for low and high flux. In a simplified analysis one can evaluate these regions separately. The slopes are close to 1 and 2 indicative of the expected behavior. The intercept of those lines occurs at a primary flux at which the number of `true'\ and `random'\ coincidences are equal. We will present below an improved procedure to  analyze the data.
The key point is that for this study it is proven that`true'\ coincidence exist. Furthermore, one can adjust the primary flux such that the ratio of true-to-random coincidences (termed TR ratio) is at an acceptable level. The usual wisdom is to aim for a value above 1.
This type of analysis can be used for energy dispersive experiments and ToF spectrometer. Implicit is the assumption that the primary flux can be increased to such a high level that `random'\ coincidences dominate the count rate. While this is in many cases not a constraint it can be an issue. In such a  case the log-log plot can not be generated.\linebreak
\indent In the pioneering experiment of Bothe and Geiger  a different way appropriate for energy dispersive experiments was discussed.\cite{1389_Bothe} It is based on the measurement of the arrival time difference of the two emitted electrons at their respective detectors, see Fig.\ref{hist_cartoon}.
The implementation for an instrument which consists of a pair of hemispherical analyzers has been discussed in detail, here we recall only the conceptional points. \cite{1537_vanRiesen,1425_vanRiessen,1588_Schumann}
For a valid coincidence event the arrival time ($t_{left}$ and $t_{right}$) of two electrons at the respective detector with respect to the coincidence trigger is known. Therefore we can plot the coincidence intensity as a function of the time difference $dt$=$t_{left}-t_{right}$. A schematic curve is shown in Fig.\ref{hist_cartoon}.
The emergence of a peak is evidence of `true'\ coincidences as discussed in the literature.\cite{1389_Bothe,1098_Jensen,1624_Amaldi,1494_McCarthy,1696_Hayes, 1039_Thurgate,888_Kirschner}. There is no temporal relation between the `random'\ coincidences which explains the constant intensity outside the peak region. Apart from  proving the existence of  `true'\ coincidences this approach allows the straightforward determination of the TR ratio of `true'\ to `random'\ events.
\indent The total width of the $dt$ curve is given by $t_{tot}$. The base width of the peak is   $t_{true}$ as indicated by the pair of vertical dashed lines, see Fig.\ref{hist_cartoon}. This result is used to define a region of interest where `true'\ coincidences are located.  The peak width  is in accordance with the time dispersion of the spectrometer.\cite{1493_Volkel,1479_Imhof, 1638_Kugeler} \linebreak
%
%
\indent The generation of a log-log plot with suitable statistics takes a considerable time and will interrupt the actual measurement, because largely different primary fluxes have to be used.\linebreak  
\indent In this respect the use of a histogram is very powerful. We can illustrate this point by  observing how the histogram evolves. For this we take a data file from a DPE experiment from a Pb surface. The experiment in question captured the coincidence of a $5d$ core electron and related Auger electron upon excitation with 40.8~eV photons, see Fig.8 of Aliaev et al.\cite{2176_Aliaev}. The coincidence rate was around 1~cps for this study.

\begin{figure}[t]
	\includegraphics[width=8.0cm]{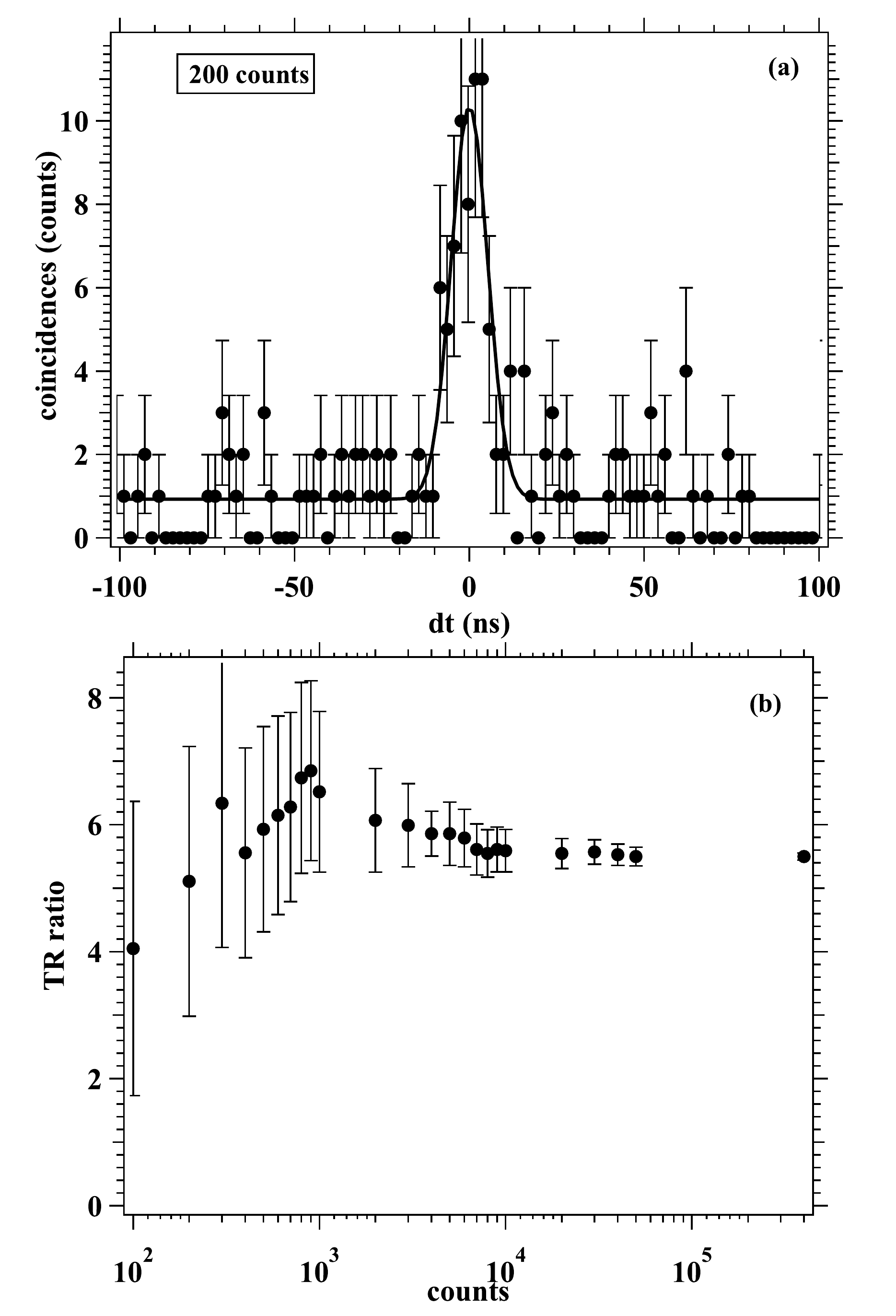}
	\caption{ Panel  (a) shows the  arrival time histograms for 200 coincidence counts. The evolution of the TR ratio as a function of the accumulated counts is plotted in panel (b). The data are from a measurement from a Pb surface, see Fig.8 (a) in Aliaev et al\cite{2176_Aliaev}.\label{Pb_hist}}
\end{figure}

The curve of Fig.\ref{Pb_hist} (a)  shows the histogram after data acquisition of 200 coincidence events which for the given coincidence rate translates to 200 s. Even after this short time it is clear that `true'\ coincidences exist and the TR ratio can be computed. The evolution of the TR ratio as a function of the counted events is displayed in Fig.\ref{Pb_hist} (b) and one can see that the value of the TR ratio is accurately known after around 2000 counts.

At the beginning of a new study the required intensity level for a sufficiently good TR ratio is not known, hence the primary flux needs to be varied. For a log-log plot the picture emerges only  after a sufficiently large flux window of 2-3 orders of magnitude, see Fig.\ref{log_log_scienta}. This is different for the histogram measurements where a single measurement gives a result.
\section{delayed coincidence for ToF}
\begin{figure}[b]
	\includegraphics[width=9.0cm]{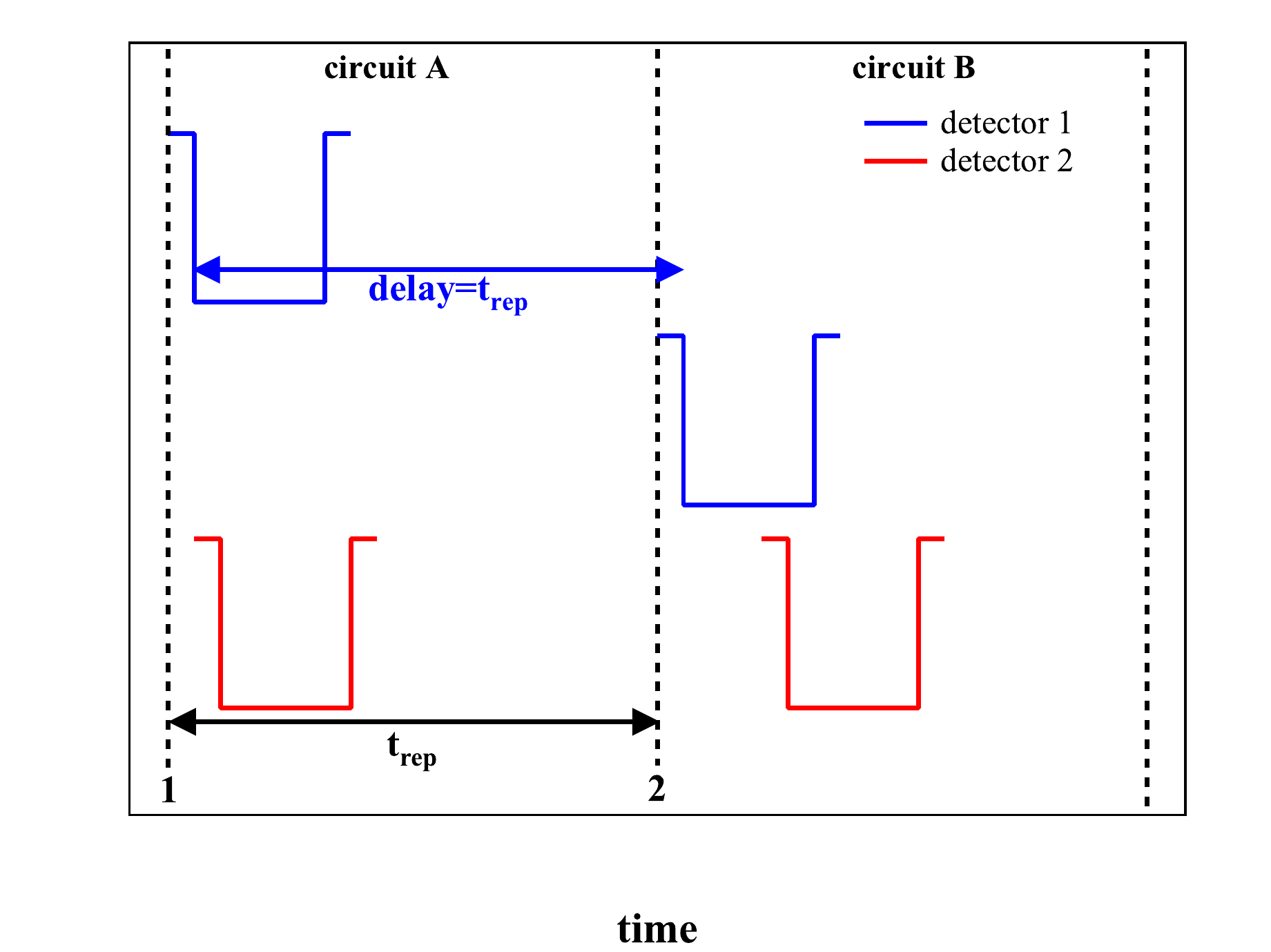}
	\caption{Schematic signals from two detectors. The dashed vertical lines indicate the time when photons from the different pulses arrive at the sample. The time between subsequent pulses is given by $t_{rep}$.  Circuit A triggers if  both detectors supply a signal after photon absorption at the first pulse and display temporal overlap. Circuit B triggers if the temporal overlap exists between the delayed signal of detector 1 and a signal from detector 2 from the next pulse. \label{cartoon_delayed}}
\end{figure}

%
%
%
%
Our ToF experiment is not set-up to record two electrons per photon pulse on one of the MCP detectors. This limitation is not as severe as it appears, because previous work has shown that the emission of two electrons into the same angular range is strongly reduced. This was related to an important concept of solid state theory known as exchange-correlation hole.\cite{1057_Schumann,1094_Schumann,1302_Hattass,1214_Schumann,1430_Schumann,2304_Schumann} The angular range of the reduced intensity region is of the order of 60$^\circ$ and exceeds the angular acceptance of the entrance lenses which is $\pm15^{\circ}$.\linebreak
\indent It is highly desirable to be able to perform an equally fast determination of the TR ratio for a ToF experiment. This becomes possible if we employ a second coincidence circuit. For simplicity and also reflecting our experimental situation we assume that the time $t_{rep}$ between two subsequent primary pulses is longer than the time-of-flight of the slowest detectable electron.
In the first circuit the two multi-channelplate (MCP) signals of the different spectrometers are used.  The second   circuit, called B, receives one MCP signal which is delayed by the time $t_{rep}$ between two consecutive primary pulses. A schematic view of the two coincidence circuits is seen in Fig.\ref{cartoon_delayed}. 
While circuit A records `true'\ and `random'\ coincidences circuit B can only record  `random'\ events. As indicated by Fig.\ref{true_random_cartoon} we define a `random'\ event  as being caused by two independent photons. This is clearly the case for circuit B. 
The probability to find two photons in the same pulse is given by $P_{2}(\lambda)$. This is in contrast to find one photon in one pulse while the second photon is in the subsequent pulse. For this scenario the probability is given by the joint product of probabilities $P_{\lambda}(1)^2$. For $\lambda\ll1$ we can simplify  $P_{\lambda}(1)^2=2\cdot P_{\lambda}(2)$ by using the expansions of Eq.(\ref{poisson1}). 

In the case of a `random'\ coincidence, circuit A will record the first electron  either by detector 1 or 2. The second electron is then registered in the opposite detector. In other words there exists two detection combinations in contrast to circuit B. This means that the `random'\ rate for both circuits is the same and we can write:
\begin{equation} c_A=r+t \\, \quad   c_B=r
	\label{counter}
\end{equation}
The coincidence rates for circuits A and B are given by $c_A$ and $c_B$, while the terms $r$ and $t$ refer to the `random'\ and `true'\ coincidence rates. This terms are low count rate approximations as discussed in the literature.\cite{2371_Kossmann} With this we define $ratio$ as:
\begin{equation} ratio=\frac{c_A}{c_B}=\frac{r+t}{r}=1+\frac{t}{r} 
	\label{counter_ratio}
\end{equation}
We immediately see that $ratio$ is closely related to the quantity of interest, namely the TR ratio. We recall that the `random'\ rate varies quadratic with flux while the `true'\ events scale linearly with the flux. Since the singles rate also depends linearly on the flux, a presentation of $ratio$ as a function of the inverse singles rate is expected to yield a line with an intercept of 1. If we label the singles rate with $s$ we can formulate:
\begin{equation} ratio=a_1+a_2 \cdot \frac{1}{s}
	\label{counter_ratio2}
\end{equation}
\begin{figure}[b]
	\includegraphics[width=7.5cm]{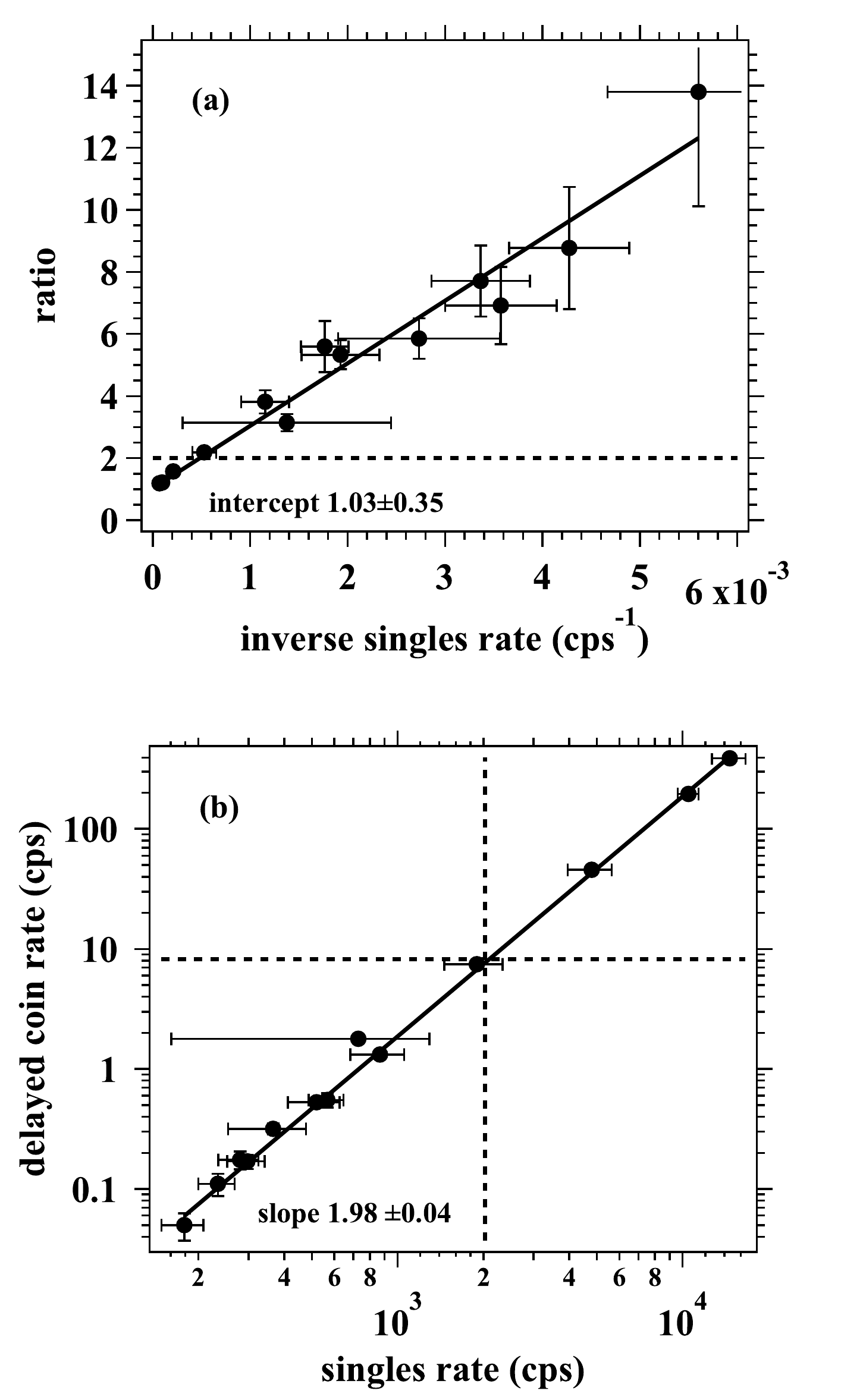}
	\caption{ In (a) the measurement of the $ratio$ as a function of inverse singles rate  is shown. These are  DPE data come from a SrRuO$_3$ surface with a photon energy of 30 eV. The solid line is  a linear fit to the data and reveals an intercept $a_1$=1.03$\pm 0.35$. The horizontal dashed line indicates $ratio=2$ which is equivalent to TR=1. In panel (b) we plot the delayed coincidence rate as a function of the singles rate via a log-log plot. We find for the slope $b_2$=1.98$\pm 0.04$. \label{SrRuO_30eV}}
\end{figure}
We expect the intercept $a_1$ to be close to 1 according Eq.(\ref{counter_ratio}). In this case the slope $a_2$ can be identified with the singles rate $s$ for which ratio is close to 2 or equivalently  the TR ratio is 1.\linebreak
\indent A single photon pulse will lead with a probability $p_s$ to the detection of an electron in one spectrometer. The time between subsequent pulses is given by $t_{rep}$. Therefore the single electron rate is :
\begin{equation} s=\frac{p_s}{t_{rep}}
	\label{singles}
\end{equation}

Similarly we get for the delayed coincidence rate $c_b$:
\begin{equation} c_b=\frac{p_s^2}{t_{rep}}=s^2 \cdot t_{rep}
	\label{delayed_rate}
\end{equation}
If we take the logarithm of both sides we obtain: 
\begin{equation} log(c_b)=log(t_{rep})+2\cdot log(s)=b_1+b_2\cdot log(s)
	\label{delayed_rate}
\end{equation}
This means that the fitting of the data in a double logarithmic fashion will yield fit parameters $b_1$ and $b_2$. For our experiment with a $t_{rep}=2 \cdot 10^{-6}$ sec we expect $b_1$=-5.699 and $b_2$=2. 
In order to explore the validity of the previous statements we have performed experiments using a ToF instrument with an HHG light source as described previously.\cite{1804_Huth,2111_Truetzschler,2268_Chiang} The spectrometer settings were identical to those reported in our previous work. We selected a SrRuO$_3$ sample as target which we investigated with a  few different photon energies.\linebreak 
\indent In Fig.\ref{SrRuO_30eV} (a) we show the behavior of $ratio$ as a function of the inverse singles rate $s^{-1}$ for a photon energy of 30~eV. It is apparent that the data are well-described by a line which has an intercept of $a_1$=1.03$\pm$0.35  in agreement with the expected value of 1. The horizontal dashed line marks the level for $ratio$=2 or TR=1. For the other parameter we find $a_2$=2013.5 cps which is singles rate for which the TR ratio is near 1. In Fig.\ref{SrRuO_30eV} (b) we show the delayed coincidence rate as a function of the singles rate in double logarithmic manner. A linear fit reveals a slope of $b_2$=1.98$\pm0.04$ which is within the error of the expected value for `random'\ coincidences. The vertical dashed line marks the singles rate $s=a_2$ and the crossing point with the solid line occurs for a delayed coincidence rate of 7.87 cps which is identical to the `true'\ rate. We obtain a value of the fit parameter $b_1$=-5.6479 which is within the expected range.\linebreak
\begin{figure}[t]
	\includegraphics[width=7.5cm]{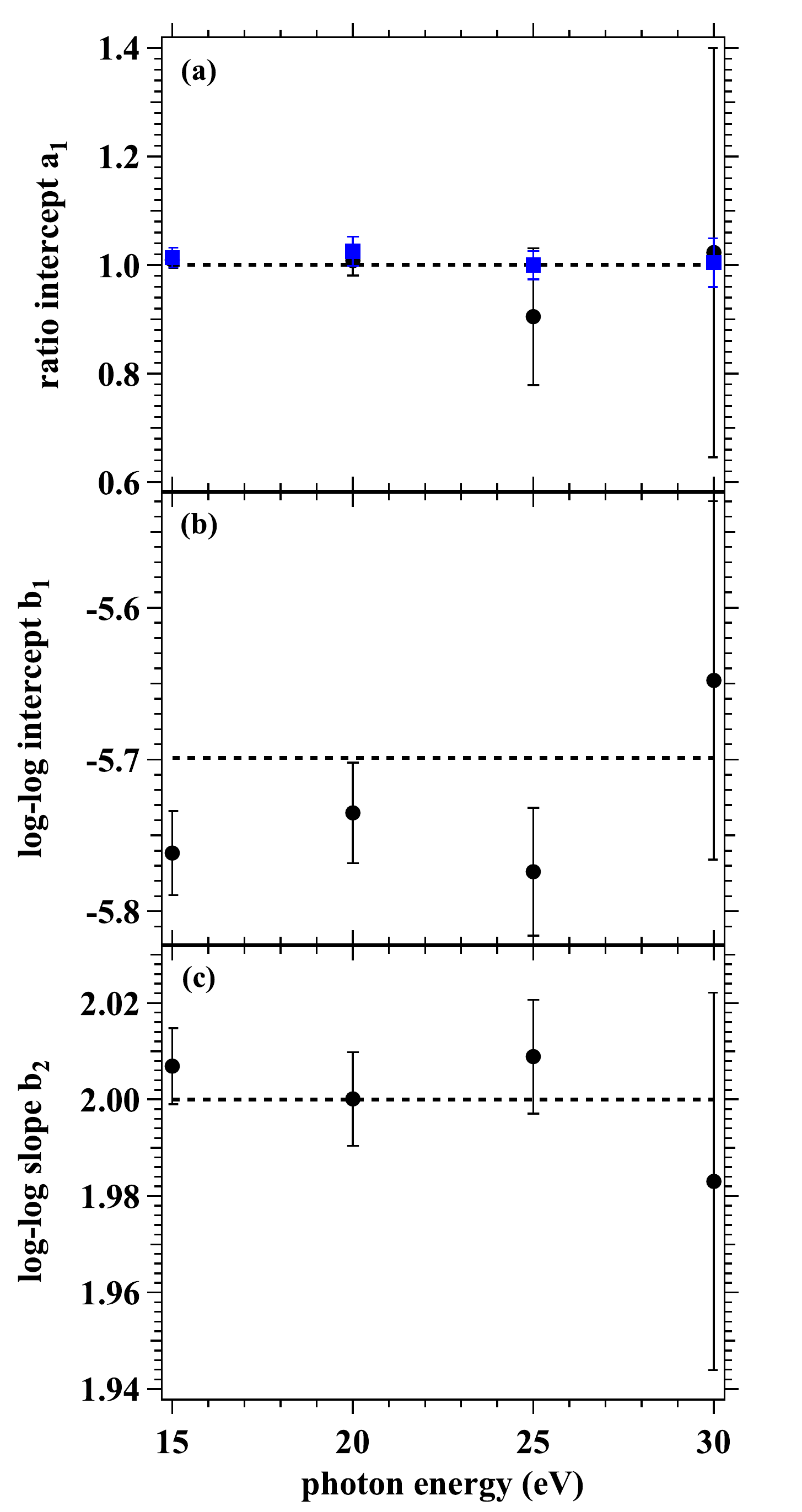}
	\caption{We summarize the flux dependent measurements for a variety of photon energies. In the   panels (a)-(c) we show the variation of the fit parameters $a_1$, $b_1$ and $b_2$ with their error bars. The dashed horizontal lines in each plot are the expected values as discussed in the text. \label{summary_delayed}}
\end{figure}
\indent We summarized our findings for a variety of photon energies in Fig.\ref{summary_delayed}. The three panels show the variation of the fit parameters $a_1$, $b_1$ and $b_2$ together with the expected values as indicated by the dashed horizontal lines in each plot. The blue data points for $a_1$ of Fig.\ref{summary_delayed}{(a) are obtained if only the data  with highest primary flux are retained. In this case $a_1$=1 is almost exactly fulfilled. The key point is that the material independent parameters are essentially in agreement with the prediction. 

\indent In the description so far we have employed 4 parameters which can be determined by fitting the experimental data. It is appropriate to perform calibration measurements with the aim to keep some parameters fixed. The intercept $a_1$ from Eq.(\ref{counter_ratio2}) can be easily measured by operating at a high singles rate. This is best done at lower photon energies, because there we observe  a lower `true'\ rate for the same singles rate. 
For the evaluation of the `true'\ and singles rate for  TR=1 we will set $a_1$=1 and $b_2$=2. In other words in our analysis we adjust only two parameters. We compare the results with the usual double logarithmic analysis of the coincidence rate $r_a$ versus the singles rate $s$ formulated as:
\begin{equation} log(r_a)=log(a \cdot s + b\cdot s^2)
	\label{loglog_eq}
\end{equation}
The parameters $a$ and $b$ refer to the contribution of `true'\ and `random'\ coincidences, respectively. Once the parameters $a$ and $b$ are obtained in the analysis it is straightforward to determine the singles and `true'\ rate for TR=1. These are given by  $s=a/b$ and $t=a^2/b$. The outcome of this analysis and comparison with the result from the delayed coincidence is plotted  in Fig.\ref{delay_loglog}. It is clearly shown that both ways of analysis give essentially the same result. Furthermore, the variation of the `true'\ rate covers more than 2 orders of magnitude and starts at a low rate of 0.02 cps for a photon energy of 15 eV.  
The demonstration of consistent results clearly proves that the delayed coincidence is a valid approach. Therefore, the evaluation of the TR ratio without the need of a flux variation is possible. Hence monitoring the TR ratio during a measurement similar to those for the energy dispersive set-up shown in Fig.\ref{Pb_hist} is possible.
\begin{figure}[t]
	\includegraphics[width=8.5cm]{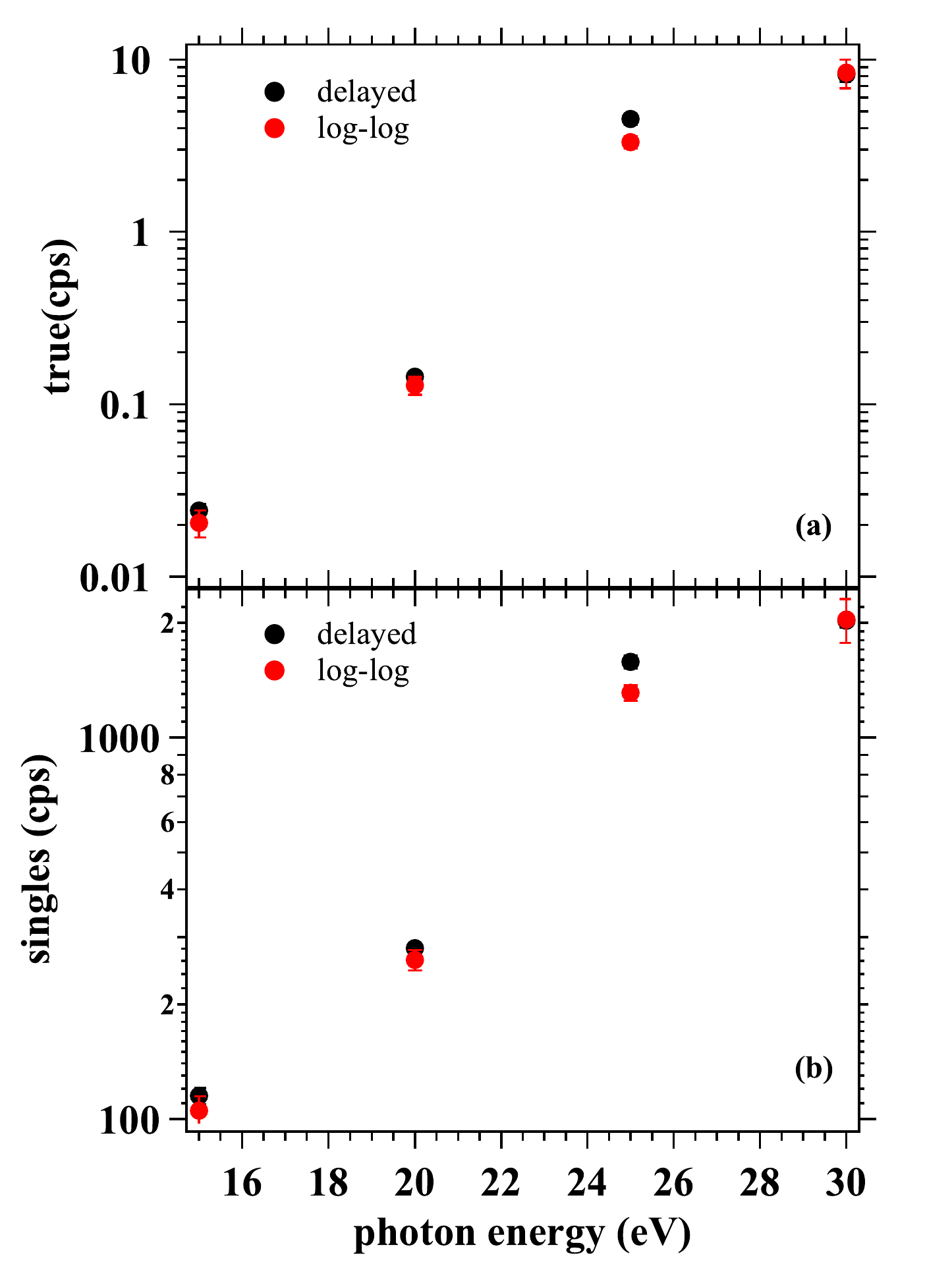}
	\caption{Photon energy dependence of the `true'\ (a) and singles (b) count rate for TR=1. The black data points refer to the result from the delayed coincidences while the red data points stem from a double-logarithmic analysis. \label{delay_loglog}}
\end{figure}
\section{Removal of `random'\ background}
Besides the knowledge of the `random'\ contribution to the count rate it is also beneficial to be able to remove the `random'\ background from the energy spectra without the need to collect an additional spectrum at very high primary flux. This possibility exists for the energy dispersive instrument and we have  discussed our implementation in the literature.\cite{1588_Schumann,2250_Schumann} 
It turns out that this can also be achieved with a ToF instrument as shown  in a recent work on APECS at a synchrotron light source.\cite{2325_Leitner} Effectively the concept of the delayed coincidence was used which was facilitated by the possibility to record the absolute time for each photon bunch.
One can make use of a delayed coincidence circuit to get the equivalent information.
The key point is to change the operation of the coincidence circuit and to add an additional channel in which the signal level indicates the presence of a delayed coincidence. Rather than using the AND operation with the two MCP signals one selects the OR option. Therefore, each time at least one spectrometer records an electron impact a trigger is generated. At this point the detector signals are read and stored in a list as for the usual AND operation. In Table \ref{table} we provide the schematic data structure. The 1st column is the recorded event number. The 2nd and 3rd columns indicate whether the respective spectrometer, labeled as $left$ and $right$, has kinetic energy and emission angle information. Consequently the entry is either $yes$  or $no$ for each of these two columns. 
\begin{table}[b]
	\caption{\label{tab:example}Schematic data structure.}
	\begin{ruledtabular}
		\begin{tabular}{ccccc}
			event  & left spect & right spect & coin & delayed coin\\
			\hline
			1 & yes   & yes & yes & no \\
			2 & yes   & no   & no &no \\
			3 & no    & yes  & no &yes \\
			4 & no    & yes  & no & no \\
			5 & yes   & no  & no &no \\
			6 & no    & yes  &no &yes \\
			.. & ..    & ..  & .. &.. \\
		\end{tabular}
	\end{ruledtabular}
	\label{table}
\end{table}
If for a given event both spectrometer registered an electron this is a coincidence event which gives the entry $yes$ in the column $coin$. This is the case for the 1st event in Table \ref{table}. This means that the usual AND condition of the coincidence circuit can be recovered in a post-experiment analysis and the energy spectrum containing both `true'\ and `random'\ coincidences can be computed.\linebreak 
\indent Alongside the coincidence circuit which was changed to an OR operation we will operate a delayed coincidence circuit in the AND mode as sketched in Fig.\ref{cartoon_delayed} as circuit B. This provides the signal for the last column of the table. It has an entry if a delayed coincidence circuit  has triggered. This is the case for event 3 where only the $right$ spectrometer records an electron. However, the  $left$ spectrometer has an entry for event 2. This means that the events 2 and 3 originate from two photon pulses separated exactly by the time $t_{rep}$. This  means that  we can in a post-experiment analysis identify a delayed `random'\ coincidence and compute the energy spectrum. If we remove this spectrum from the one containing `true'\ and `random'\ we obtain effectively the `true'\ spectrum. This procedure is allowed, because we have established above that the number of `random'\ coincidence in the regular coincidence circuit is identical to those in the delayed circuit. 
 \section{revised definition of $\lambda$}
 In the discussion so far we have introduced the primary intensity via the average number of photons per pulse $\lambda$. However, we are talking about electron detection, hence we have to introduce a parameter which describes the probability for electron emission upon photon absorption and detection within a given spectrometer. This will be a material and photon energy dependent entity which we want to abbreviate  with $y$ for yield.  The focus are `random'\ events which are mainly determined by single electron emission, hence we ignore double electron emission at this moment. The value of $y$ will depend  also on the detailed spectrometer settings which determines the solid angle of detection and the probed energy window. This will lead to a singles count rate $s$ which is easily determined. Conceptually the photon flux can be measured in absolute units and with the known time between pulses $t_{rep}$ the relation between $\lambda$ and $y$ reads as:
 \begin{equation} y=\frac{s\cdot t_{rep}}{\lambda} 
 	\label{def_y}
 \end{equation}
 This means that the average number of detected electrons per pulse is simply $y \cdot \lambda$. In a  next step one would like to know the probability function for emitted electrons. In other words what is the probability that $m$ electrons are detected. For this evaluation one needs first the distribution function for $n$ photons in a pulse, which we have abbreviated with $P_n(\lambda)$ in Eq.(\ref{poisson1}). 
  Further, we do not assume that  $\lambda\ll1$. If $n$ photons are absorbed by the sample, the number of electrons $m$ registered by one spectrometer is  $m\leq n$. The probability for this to occur is determined by the binomial distribution:
 \begin{equation} el_{n,m}(y)=  \binom{n}{m} \cdot y^m \cdot (1-y)^{n-m}    
 	\label{binomial}
 \end{equation}
 This finally leads to the probability for the detection of $m$ electrons:
 \begin{equation}    P_m^{elec}(\lambda,y)=\sum_{i=1}^{n} P_i(\lambda) \cdot  el_{n,m}(y)
 	\label{binomia2l}
 \end{equation}
If the photon distribution is a Poisson distribution then one can show that $P_m^{elec}$ is also a Poisson distribution with a new parameter $\lambda_{eff}=\lambda \cdot y$. Hence we need to demand that   $\lambda_{eff}\ll 1$ in order to derive the equivalent expressions of Eq.(\ref{poisson1}).
From Fig.\ref{SrRuO_30eV} we can see that the highest singles rate was 15000 cps. With $t_{rep}$=2000 ns we obtain for $\lambda_{eff}=$0.03. Clearly the condition  $\lambda_{eff}\ll 1$ is fulfilled in our DPE measurements and the approximations made in Eq.(\ref{poisson1}) are valid.\linebreak
\indent In the discussion of the delayed coincidence we have ignored the possibility that detector 1 has a `true'\ counter part in detector 2. This will lead to additional events, however the `true'\ coincidence rate is more than 2 orders of magnitude smaller than the singles rate, see Fig.\ref{delay_loglog}. Hence this contribution can be safely ignored.\linebreak
%
\indent This observation is not tied to the current DPE measurement on SrRuO$_3$, but is a rather general result in our activities on surfaces. This includes also observations from Auger-Photoelectron coincidences. This in turn means that the consideration of $y$ to contain only single electron events is justified. This should not be read as an indication that electron pair emission from surfaces is a weak effect. A combined (e,2e) and DPE study revealed that 15-40$\%$ of electrons detected by one spectrometer have a counterpart emitted somewhere in the halfspace.\cite{2021_Schumann}
\section{Comparison energy dispersive versus ToF performance}
A variety of coincidence spectrometer have been developed addressing different scientific questions. There is no rule which determines the best approach. Here we only want to provide a simple estimate on the performance relation between an energy dispersive and ToF approach. For simplicity we assume that the emission of pairs is isotropic and constant within an energy window. We assume at the moment that the energy dispersive and ToF instrument cover the same solid angle $\Omega$ and energy window.  Later we will discuss the deviations from this description.\linebreak
\indent An example  resembling this situation is a pair of hemispherical analyzers compared to a pair of ToF spectrometers which are based on the entrance lens of a hemispherical analyzer. In our activities these approaches have been realized.\cite{1335_vanRiessen,1588_Schumann,1804_Huth,2111_Truetzschler,2268_Chiang}
The key number characterizing the performance of the hemispheres is the time dispersion. This is essentially the quantity $t_{true}$, see Figs.\ref{hist_cartoon} and \ref{Pb_hist}.  
A ToF instrument requires a pulsed excitation source and the time between subsequent pulses is $t_{rep}$.  Let us assume that the average number of primary particles per pulse is given by $\alpha$ which yields the desired ratio of `true'\ to `random'\ . This means that the primary flux for the ToF experiment is:
\begin{equation} f_{ToF}=\frac{\alpha}{t_{rep}}
	\label{int_tof}
\end{equation}
For a proper comparison with an energy dispersive instrument we have to allow the same number of primary particles within the coincidence window. The temporal width of this is characterized by the value of $t_{true}$, see Fig.\ref{hist_cartoon}.
This will ensure the same ratio of `true'\ to `random'\ and we obtain for the primary flux $f_{disp}$:
\begin{equation} f_{disp}=\frac{\alpha}{t_{true}}
	\label{int_tof}
\end{equation}
This allows now to compare the primary flux of the two types of experiments and we obtain:
\begin{equation} \frac{f_{disp}}{f_{ToF}}=\frac{t_{trep}}{t_{true}}
	\label{flux_relation}
\end{equation}

Let us use some typical values for our instruments which are $t_{true}$=20~ns and $t_{rep}$=1000~ns. This tells us that the energy dispersive system is a factor of 50 more efficient or equivalently a higher coincidence rate by this factor is possible.\linebreak
\indent However, the assumption of the same solid angle for both types of spectrometer is not valid. While our ToF instrument captures 3$\%$ of the half sphere this value is reduced to 0.7$\%$ for the hemisphere set-up. Since the solid angle enters quadratic into the detection efficiency it reduces the performance advantage to about 3.\linebreak 
\indent The energy window captured by the ToF and energy dispersive instrument can be selected. The requirement of the operation with a low primary flux makes asks for efficient detection hence a sizable window is required at the expense of energy resolution. In the context of this work we operated the hemispherical analyzers with a pass energy of 150~eV which provides an energy window of 13.5~eV. The typical settings of the ToF instrument for coincidence measurements have been described.\cite{2194_Huth} It is possible to cover an energy window of 30 eV, but only over a range of 10~eV the spectrometer captures electrons with the  maximum solid angle of 3$\%$. Outside this range the solid angle drops significantly. This means that the detection efficiency for a  ToF is about a factor of 1.25 larger than for the hemisphere. For a coincidence experiment one has to consider the square of this value and this gives a factor 1.56. From this one concludes that the performance advantage of the hemisphere reduces to a factor 2.
In this  estimate other experimental considerations have been ignored, but the main conclusion is that both approaches provide  performance on equal terms. 
\section{summary}
%
For coincidence experiment it is vital to know the ratio of `true'\ to `random'\ events which is sensitive to the primary flux.  We reviewed the established procedures on how to identify `true'\ electron pair emission. A powerful tool known for energy dispersive coincidence experiments was found to have its counter part in ToF instrumentation. This approach we termed delayed coincidence circuit. We discussed the relevant parameters of this procedure and how they can be determined in an experiment. In order to support this notion we performed photon energy dependent DPE studies. These resulted in intensity levels varying by more than 2 order of magnitude. The characteristic values of the delayed coincidence are in agreement with the predicted behavior and with the concurrent obtained double-logarithmic presentation.\linebreak
\indent This experimental proof of the concept allows to monitor the TR ratio during the actual measurement. Therefore there is no need to  resort to the time consuming flux dependency measurement for the conventional double-logarithmic analysis which interrupts the actual measurement. Furthermore, it is not required to have such a high primary flux in which the coincidence rate is dominated by `random'\ coincidences.\linebreak
\indent We have also outlined how it becomes possible to remove the `random'\ contribution from the spectra by using the novel approach.  
%
 
\bibliographystyle{aip}

\end{document}